\documentclass[prb,aps,showpacs,amsmath]{revtex4} 
\usepackage{amssymb}
\usepackage{latexsym}
\usepackage{graphicx}

\usepackage{epsfig}

\usepackage{dcolumn}
\usepackage{amsmath}
\usepackage{amsfonts}
\usepackage{bm}

\newcommand{\be}{\begin{equation}}
\newcommand{\ee}{\end{equation}}
\newcommand{\bea}{\begin{eqnarray}}
\newcommand{\eea}{\end{eqnarray}}

\newcommand{\p}{\partial}
\newcommand{\s}{\sigma}

\newcommand{\la}{\langle}
\newcommand{\ra}{\rangle}
\newcommand{\rd}{\mbox{d}}
\newcommand{\ri}{\mbox{i}}
\newcommand{\re}{\mbox{e}}

\renewcommand{\vec}[1]{{\bm #1}}

\begin{document}
\title{ Two-leg SU($2n$) spin ladder:  A low-energy effective field theory approach}
\author{P. Lecheminant$^1$ and A. M. Tsvelik$^2$}
\affiliation{$^1$Laboratoire de Physique Th\'eorique et
Mod\'elisation, CNRS UMR 8089,
Universit\'e de Cergy-Pontoise, Site de Saint-Martin,
2 avenue Adolphe Chauvin,
95302 Cergy-Pontoise Cedex, France.\\
$^2$Department of Condensed Matter Physics and Materials Science, Brookhaven National Laboratory, Upton, NY 11973-5000, USA}
 \date{\today } 

\begin{abstract} 
We present a field theory analysis of a model of two SU($2n$)-invariant magnetic chains coupled by a generic interaction preserving time reversal and inversion symmetry. Contrary to the SU(2)-invariant case the zero-temperature 
phase diagram of such two-leg spin ladder does not contain topological phases.  Only generalized Valence Bond Solid 
phases are stabilized when $n>1$ with different wave vectors and ground-state degeneracies. In particular,  
we find a phase which is made of a cluster of $2n$ spins put in an SU($2n$) singlet state. For $n=3$, this cluster phase
is relevant to $^{173}$Yb ultracold atoms, with an emergent SU(6) symmetry, loaded in double-well optical lattice.

\end{abstract}

, 

\maketitle
\section{Introduction}

 Interacting quantum systems with symmetries higher than the ubiquitous SU(2) one may display 
 new interesting physics. \cite{wuzhang}
 For instance, chiral spin liquids with non-Abelian statistics might emerge in quantum magnets
 with an extended SU($N$) symmetry.\cite{gurarie}
 Although high symmetries are more difficult to maintain, there are several interesting possibilities of their realization. For instance, in their recent preprint Kugel {\it et.al.} \cite{khomskii} suggest condensed matter realizations of the antiferromagnetic chains with high symmetry, such as SU(4) in the context of orbital degeneracy. \cite{li,ueda,milasun,pati,azaria}
The second possibility is related to alkaline-earth or ytterbium ultracold atoms which have 
a peculiar energy spectrum.  The ground state and a metastable first excited state have zero electronic angular momentum 
so that the nuclear spin $I$ is almost decoupled from the electronic spin. 
This decoupling paves the way to the experimental realization of magnets with
an emergent SU($N$) symmetry where $N= 2 I +1$  is the number of nuclear states which
can be as large as 10 with strontium and ytterbium atoms.\cite{gorshkov,Cazalilla-H-U-09,Cazalilla-R-14}
Recent experiments with such atoms loading into optical lattices have directly observed the existence of the SU($N$) symmetry and have determined the specific form of the interactions
between atoms in the ground state and the excited state. \cite{Scazza-et-al-14,Zhang-et-al-14} 

The simplest lattice SU($N$) model stems from alkaline-earth atoms in their ground state loaded into the lowest band of an 1D optical lattice with a filling of one atom per site which best avoids three-body looses. For large repulsive interaction, 
the resulting SU($N$) symmetric magnet is described by the Hamiltonian:
\bea
{\cal H} = J\sum_{n} \hat P_{n,n+1}, \label{single}
\eea
where $\hat P_{n,n+1} = \sum_{A=1}^{N^2-1} S^A_n S^A_{n+1} + $const, with $S^A$ being generators of the su($N$) Lie algebra, is a permutation operator acting in the tensor product of the $N$-dimensional Hilbert spaces $n$ and $n+1$. Model (\ref{single}) is integrable by means of the Bethe-Ansatz approach. \cite{sutherland} For $J >0$ its excitation spectrum is gapless and at small energies the dispersion is linear. Hence the theory is critical with $N-1$ gapless relativistic modes; 
its universality class is the SU($N$)$_1$ Wess-Zumino-Novikov-Witten (WZNW) model 
with central charge $c=N-1$. \cite{affleckspinchain,affleck} 

When one considers chains coupled by generic interactions the integrability is lost though at the same time the physics becomes more interesting, as one can learn from the example of the problem of two coupled $S=1/2$ Heisenberg chains. \cite{dagotto,shelton,giamarchi,bookboso,Giamarchi-book} The latter model has a phase diagram which includes topological phases containing seeds of superconductivity. \cite{fabrizio, schulz,fisher,rvb1,rvb2,shura}
 It also may used to describe experimentally observed confinement of fractional quantum number excitations  existing for the single chain problem. \cite{rvb2,lake}

 In this paper  we consider a one-dimensional version of the famous Kugel-Khomskii model, \cite{kugel} namely a model of two weakly coupled SU($N$) chains (labeled $1,2$):
\bea
{\cal H} = 
\sum_n\Big[\hat P_{n,n+1}^{(1,1)} + \hat P_{n,n+1}^{(2,2)} + \lambda \hat V^{(1,2)}_{n,n}\Big], \label{ladder}
\eea
where $\hat P^{(a,a)}$ are permutation operators acting on states on chain number $a=1,2$. In the simplest case the interaction $\hat V$ is just a permutation operator acting between sites of different chains. However, to select  for our analysis any specific form of the interaction would be too restrictive. Instead we will use an alternative approach, namely, we will consider weakly coupled SU($N$) chains with general interaction, which may go beyond the two-spin exchange. As the result, the problem becomes then a perturbed conformal field theory (CFT) with Hamiltonian:
\bea
{\cal H} = W[SU(N)_1]_1 +W[SU(N)_1]_2 
+ \sum_{a,b} \lambda_{ab}  \int dx \; {\cal O}_{1,a}(x){\cal O}_{2,b}(x), \label{perturbed}
\eea
where $W[SU(N)_1]_{1,2}$ corresponds to the WZNW Hamiltonian for SU($N$)$_1$ CFT on the corresponding chain 
and ${\cal O}_{a,1}$ and ${\cal O}_{2,b}$ are  operators acting on states of chains 1,2. In the weak coupling limit 
$|\lambda_{ab}| << J$ one may take into account only relevant perturbations which greatly restricts the number of possible operators. On top of the SU($N$) symmetry, our choice will be further restricted by considering only (i) spatially homogeneous ladders and (ii) ladders with inversion symmetry. The latter consideration excludes operators with non-zero conformal spin. 
In this way a multidimensional phase diagram of the lattice model is projected on a manageable (in fact, in most cases just two-dimensional) phase diagram of the perturbed CFT. 

We will investigate the infrared (IR) physics of model (\ref{perturbed}) when $N=2n$. Our main conclusions are that the physics of the SU(2$n$) ladder with $n >1$ in some crucial aspects is different from the SU(2) case. In particular, there are no topological non-degenerate phases. All phases contain local order parameters corresponding to Valence Bond Solids (VBS). There are crystals of two types: with wave vectors $(\pi/n, 0)$ or $(\pi/n,\pi)$ ($2k_F$ VBS) and $(2\pi/n, 0)$  ($4k_F$ VBS). The latter possibility did not exist for the spin $S=1/2$ ladder. It can be viewed as a cluster of $2n$ spins put in an SU($2n$) singlet
state. In the simplest case, i.e., $N=4$, this cluster corresponds to the plaquette phase found in the numerical
analysis of Ref. \onlinecite{mila} of the SU(4) two-leg spin ladder with an antiferromagnetic interchain coupling.
For $N=6$, we  find the emergence of a cluster phase of six spins, leading to trimerization which should occur in the phase
diagram of the two-leg SU(6) spin ladder. The latter case is directly relevant to the insulating phase
of double tube of ytterbium $^{173}$Yb ultracold atoms.

The rest of the paper is organized as follows. In Sec. II, the continuum limit of weakly-coupled SU($2n$) two-leg spin ladder is presented to identify the leading perturbation in Eq. (\ref{perturbed}). The result is the continuum model (\ref{cft}). In Sec. III this model is analyzed using  the conformal embedding approach. As a result model (\ref{cft}) is expressed as a theory of 
${\mathbb{Z}}_N$ parafermions coupled to SU($N$)$_2$ WZNW via some relevant operator (\ref{cftnewbasis}). This new formulation allows us to determine the nature of the phases. The resulting analysis is provided in Sec. IV and our concluding
remarks are given in Sec. V. Finally, our paper is supplied with several appendices where some additional technical
details are described.

\section{The continuum limit}
In this section, we determine the leading perturbation of model (\ref{perturbed}) by means of a continuous description
of two-leg SU($2n$) spin ladder with generic interactions.

Let us start with the decoupling limit where the lattice model (\ref{ladder}) reduces to two decoupled Sutherland models (\ref{single}). Its low-energy properties can be obtained by starting from
the U($N$) Hubbard model at $1/N$ filling with large repulsive $U$ interaction.  \cite{affleckspinchain,affleck,assaraf,manmana}
At energies below the charge gap $\sim U$, the SU($N$) spin operators in the continuum limit are
described by: \cite{affleckspinchain,affleck,assaraf}
\begin{equation}
S^{A}_{l} \simeq J^{A}_{ lL}(x) +  J^{A}_{lR}(x) + \re^{ \ri 2k_F x} N^{A}_l(x) + \mbox{e}^{-\ri 2k_F x} N^{A \dagger}_l(x)
+  \mbox{e}^{ \ri 4k_F x} n^{A}_l(x) + ..,
\label{spinop}
\end{equation}
where $l=1,2$ denotes the two decoupled chains and $k_F = \pi/Na_0$ ($a_0$ being the lattice spacing)
since the underlying Hubbard model is $1/N$ filled (1 electron per site).
In Eq. (\ref{spinop}), $J^{A}_{l L,R}$ are the left and right currents which generate 
the SU($N$)$_1$ CFT. They are defined in terms of the underlying left- and right moving Dirac fermions 
$L_{l\alpha}, R_{l\alpha}$ as 
\begin{equation}
J^A_{l R} = R_{l \alpha}^{\dagger} T^A_{\alpha \beta}
R_{l \beta} , \; \;  
J^A_{l L} = L_{l \alpha}^{\dagger} T^A_{\alpha \beta}
L_{l \beta} ,
\label{suNcur}
\end{equation}
where a summation over repeated greek indices (SU($N$) indices) is implied.
In Eq. (\ref{suNcur}), $T^A$ is a generator of the su($N$) Lie algebra in the fundamental representation, 
the $2k_F$ and $4k_F$ parts of the spin density are related to primary fields of the SU($N$)$_1$ WZNW model. They 
transform respectively in the $N$-dimensional fundamental and the $N(N-1)/2$-dimensional antisymmetric representation  of 
the SU($N$) group. In particular, we have 
\begin{eqnarray}
N^{A}_l = \lambda   \; {\rm Tr} ( g_l  T^A),
 \label{2kf}
\end{eqnarray}
$ \lambda$ being a constant, related to the charge degrees of freedom, 
which can be chosen real for a matter of convenience. In Eq. (\ref{2kf}),
$g_l$ is the  SU($N$)$_1$ primary field, or WZNW field, for 
the $l$ th chain with scaling dimension $(N - 1)/N$.
This operator transforms in the $N$-dimensional fundamental representation of SU($N$) 
and can be expressed in terms of the fermionic operators through the non-Abelian bosonization approach:\cite{knizhnik,witten,affleckspinchain}
\begin{eqnarray}
g_{l \beta \alpha} \sim  \mbox{e}^{-i\sqrt{4 \pi/N} \Phi_{lc}}  L_{l \alpha}^{\dagger}  R_{l \beta},
 \label{grep}
\end{eqnarray}
$\Phi_{lc}$ being a bosonic field which captures the properties of the charge degrees of freedom
of each chain $l=1,2$.

From the continuum representation (\ref{spinop}),  we derive  a transformation of the SU($N$)$_1$ 
WZNW fields with respect to the one-step translation symmetry $T_{a_0}$:
 \begin{equation}
T_{a_0}: \; \; g_{1,2 } \rightarrow \re^{2 i \pi /N} g_{1,2}  .
\label{trans}
\end{equation}

The next step of the approach is to find the leading perturbation in Eq. (\ref{perturbed}).
One can identify it by means a of a symmetry analysis. To this end,
let us recall what symmetry restrictions we adopt. First, we have the SU($N$) symmetry:
 $g_l \rightarrow U g_l  U^{\dagger}$, with $U$ belonging to SU($N$). 
Second, we consider the spatially uniform model. Then since matrix operator $g_l$ is not invariant with respect to translations (\ref{trans}) the perturbation can include only products of $g_1g_2^+$ (or $g_1^+g_2$). From the inversion symmetry it follows that there are no operators with nonzero conformal spin such as, for example, Tr$g_1^+\p_x g_2$.  These considerations yield the following Hamiltonian: 
\bea
{\cal H} = W[SU(N)_1;g_1] + W[SU(N)_1;g_2] + \lambda_1
 \int dx \; \Big[\mbox{Tr}(g_1g_2^+) + H.c.\Big] + \lambda_2
  \int dx \;  \Big[\mbox{Tr}g_1\mbox{Tr}g_2^++ H.c.\Big]+..., \label{cft}
\eea
where the dots stand for less relevant operators like marginal current-current interaction. For the case when the interchain interaction includes only two spins, i.e., $\hat V^{(1,2)}_{n,n} = J_{\perp} \sum_A  S^{A}_{1,n} S^{A}_{2,n}$, one can check directly the form of Eq. (\ref{cft}). By substituting Eqs. (\ref{spinop},\ref{2kf}) into the interaction term of Eq. (\ref{ladder}) and taking into account that the individual chains are described by the SU($N$)$_1$ WZNW model, we find model (\ref{cft}) with $\lambda_2 = - \lambda_1/N$ and $\lambda_1 = J_{\perp} \lambda^2 /2$. 
It is important to note that the ratio of the naive continuum limit $\lambda_2 = - \lambda_1/N$ is not universal and will be modified by higher-order contributions.
For $N=2n$ matrix $-g$ still belongs to the SU($N$) group. Then the transformation $g_1 \rightarrow -g_1$ leaves the WZNW part of (\ref{cft}) unchanged and changes the sign of the coupling constants of the perturbation.
This fact will enable us to identify the corresponding parts of the phase diagram of model (\ref{cft}).

\section{Conformal embedding approach}

In the weak-coupling regime, two-leg SU($N$) spin ladders are thus described by a model of two coupled SU($N$)$_1$ WZNW models perturbed by two strongly relevant perturbations with scaling dimension $2(N - 1)/N$.
The phase diagram results thus from the competition between these two terms.
Though they have the same scaling dimension, the two perturbations are of very different nature.
The one with coupling constant $\lambda_1$ is invariant under an SU($N$)$_{\rm L}$ $\times$ 
SU($N$)$_{\rm R}$ symmetry: $g_l \rightarrow U_{\rm L} g_l  U_{\rm R}$, $U_{\rm L,R}$ being two independent
SU($N$) matrices. In stark contrast, the second with  coupling constant $\lambda_2$  is only 
SU($N$) invariant. 
It turns out that at $\lambda_2 =0$, as it will be seen, theory (\ref{cft}) is integrable and therefore it makes sense to consider this model at $|\lambda_1| >> |\lambda_2|$ and treat the $\lambda_2$-term  as a perturbation. 

\subsection{Fateev model: $\lambda_2 =0$ case}

In this respect, we now use the following conformal embedding which singles out the SU($N$) symmetry:
\begin{equation}
SU(N)_1 \times SU(N)_1 \sim SU(N)_2 \times  \mathbb{Z}_N ,
\label{embedding}
\end{equation}
where the SU($N$)$_2$ CFT has central charge $c = 2 (N^2 -1)/(N+2)$ and 
${\mathbb{Z}}_N$ is the parafermionic CFT with central charge $c= 2(N-1)/(N+2)$. \cite{para,gepner}
The latter captures the universal properties of the critical point of the  ${\mathbb{Z}}_N$ generalization of two-dimensional (2D) Ising models. The low-temperature and high-temperature phases are described respectively 
by order and disorder parameters $\sigma_k$ and $\mu_k$ ($k=1,..,N-1$) which
 are dual to each other under the Kramers-Wannier (KW) transformation.  The latter maps the
${\mathbb{Z}}_N$ symmetry, spontaneously broken in the low-temperature phase
($\langle \sigma_k \rangle \ne 0$ and $\langle \mu_k \rangle = 0$),
onto a ${\tilde {\mathbb{Z}}}_N$ symmetry which is broken in the
high-temperature phase where $\langle \mu_k \rangle \ne 0$ and
$\langle \sigma_k \rangle = 0$.  
The order and disorder operators carry respectively  
a $(k,0)$ and $(0,k)$ charges under 
the ${\mathbb{Z}}_N \times {\tilde {\mathbb{Z}}}_N$ symmetry:
\begin{eqnarray}
\sigma_k &\rightarrow& \re^{2 \pi i m k /N} 
\sigma_k \; \; {\rm under} \; \; \mathbb{Z}_N, \; \sigma_k \rightarrow
\sigma_k \; \; {\rm under} \; \; {\tilde {\mathbb Z}}_N 
\nonumber \\
\mu_{k} &\rightarrow& \re^{2 \pi i m k/N} \mu_{k}
\; \; {\rm under} \; \; {\tilde {\mathbb Z}}_N,  \; \mu_k \rightarrow
\mu_k \; \; {\rm under} \; \; {\mathbb Z}_N ,
\label{chargeorderdisorder}
\end{eqnarray}
with $m=0, \ldots, N -1$.
At the critical point, the theory is
self-dual with a ${\mathbb{Z}}_N$ $\times$ ${\tilde {\mathbb{Z}}}_N$ symmetry
and  $\sigma_k, \mu_k$ become primary fields with scaling
dimension $d_k = k(N-k)/N(N+2)$. \cite{para}
The ${\mathbb{Z}}_N$ CFT is generated by chiral 
right and left parafermionic currents $\Psi_{k R,L}$ ($\Psi_{k R,L}^{\dagger} 
= \Psi_{N-k R,L}$, $k=1,\ldots, N-1$)
with scaling dimension $\Delta_k = k(N-k)/N$ 
which are the generalization of the Majorana fermions of the ${\mathbb{Z}}_2$ Ising model.
Under the ${\mathbb{Z}}_N$ $\times$ ${\tilde {\mathbb{Z}}}_N$ symmetry, 
$\Psi_{k L}$ (respectively $\Psi_{k R}$) carries
a $(k,k)$ (respectively $(k,-k)$) charge which means:
\begin{eqnarray}
\Psi_{k L,R} &\rightarrow& \re^{2 i \pi m k/N} \Psi_{k L,R} \; \; {\rm under} \; \; \mathbb{Z}_N 
\nonumber \\ 
\Psi_{k L,R} &\rightarrow& \re^{\pm 2 i \pi m k/N} \Psi_{k L,R} 
\; \; {\rm under} \; \; {\tilde {\mathbb Z}}_N  .
\label{chargepara}
\end{eqnarray}

The next step of the approach is to observe that model (\ref{cft}) at $\lambda_2=0$, 
being  SU($N$)$_{\rm L}$ $\times$ SU($N$)$_{\rm R}$ invariant, is independent of 
the SU($N$)$_2$ sector of the embedding (\ref{embedding}) but depends only on the ${\mathbb{Z}}_N$ parafermionic currents. 
One way to see that is to relate the $\lambda_1$ term of Eq.  (\ref{cft}), i.e. $V_1$,
in terms of the underlying Dirac fermions $ R_{l \alpha}, L_{l \alpha}$  of the continuum limit:
\begin{eqnarray}
 V_1 &=& - \frac{\lambda_1}{\lambda^2}
 \int dx \; \Big[ L_{1 \alpha}^{\dagger}  L_{2 \alpha} 
R_{2 \beta}^{\dagger}  R_{1 \beta} + H.c.\Big]  
\nonumber \\
&=& - \frac{\lambda_1}{\lambda^2}
 \int dx \; \Big[  j^{+}_L   j^{-}_R + H.c.\Big] ,
\label{V1paradirac}
\end{eqnarray} 
where we have introduced a chiral SU(2)$_N$ ${\vec j}_{L,R}$ current.
As shown in Ref. \onlinecite{para}, there is a free-field representation of an SU(2)$_N$ current
in terms of a bosonic field and the first ${\mathbb{Z}}_N$ parafermion current:
\begin{eqnarray}
j^{+}_L &=& \frac{\sqrt{N}}{2 \pi}  \mbox{e}^{\ri\sqrt{8 \pi/N} \Phi_{-cL}} \Psi_{1L} \nonumber \\
j^{z}_L &=& \sqrt{\frac{N}{2 \pi}} \partial_x \Phi_{-c L} ,
\label{su2npara} 
\end{eqnarray}
$\Phi_{-c}= (\Phi_{1c} - \Phi_{2c})/\sqrt{2}$ being the relative charge bosonic field. 
For the right sector, we have a similar expression:
\begin{eqnarray}
j^{+}_R &=&  - \frac{\sqrt{N}}{2 \pi}  \mbox{e}^{-\ri\sqrt{8 \pi/N} \Phi_{-cR}} \Psi^{\dagger}_{1R} \nonumber \\
j^{z}_R &=& 
\sqrt{\frac{N}{2 \pi}} \partial_x \Phi_{-c R} ,
\label{su2npararight}
\end{eqnarray}
where  the KW duality symmetry $\Psi_{1R} \rightarrow
-\Psi^{\dagger}_{1R}$ has been used for future convenience.

We thus find that the $\lambda_1$ term  in model (\ref{cft}) is directly related to an integrable perturbation 
of ${\mathbb{Z}}_N$ parafermions introduced by Fateev with euclidean action: \cite{fateev,FZ}
\begin{eqnarray}
{\cal S} = {\cal S}_{{\mathbb Z}_N} - \tilde\lambda\int \rd^2 x \left( \Psi_{1L}  \Psi_{1R}  + H.c. \right) ,
\label{intpert}
\end{eqnarray}
${\cal S}_{{\mathbb Z}_N}$ being the action of the ${\mathbb{Z}}_N$ CFT and
$\tilde\lambda= - \lambda_1 N/4\pi^2$. This perturbation is invariant under the ${\tilde {\mathbb Z}}_N$
symmetry but explicitly breaks  the ${\mathbb Z}_N$ symmetry as seen from Eq. (\ref{chargepara}).  

\subsection{SU(N)$_2$ perturbed CFT}

Our next step is to express the $\lambda_2$-term in Eq. (\ref{cft}) in the
SU($N$)$_2$ $\times$ ${\mathbb Z}_N$ basis.  The expression of the two  SU($N$)$_1$ 
WZNW fields $g_{1,2 }$ in the SU($N$)$_2$ $\times$ ${\mathbb Z}_N$ basis ($N>2$) was obtained 
in Ref. \onlinecite{griffin}.  We will justify their results from simple arguments based on symmetries.

To perform such analysis we need the representation of the ${\mathbb{Z}}_N$ $\times$ ${\tilde {\mathbb{Z}}}_N$
symmetry in terms of the underlying Dirac fermions.
 This can be done thanks to the definitions (\ref{su2npara},\ref{su2npararight}).
Since $\Psi_{1 L}$ and $\Psi_{1 R}$ have  $(1,1)$ and $(1,-1)$ charges under
the ${\mathbb{Z}}_N \times {\tilde {\mathbb{Z}}}_N$ symmetry, we find that the ${\mathbb{Z}}_N$ symmetry
is implemented as follows on the fermions: 
 \begin{equation}
L_{1 \alpha} \rightarrow \re^{- i \pi m /N} L_{1 \alpha}, \; \;
L_{2 \alpha} \rightarrow \re^{ i \pi m /N} L_{2 \alpha}, \; \;
R_{1 \alpha} \rightarrow \re^{i \pi m /N} R_{1 \alpha} , \; \;
R_{2 \alpha} \rightarrow \re^{- i \pi m /N} R_{2 \alpha} , \; \;
\label{znfer}
\end{equation}
while under ${\tilde {\mathbb Z}}_N$ we have:
\begin{equation}
L_{1 \alpha} \rightarrow \re^{- i \pi m /N} L_{1 \alpha}, \; \;
L_{2 \alpha} \rightarrow \re^{ i \pi m /N} L_{2 \alpha}, \; \;
R_{1 \alpha} \rightarrow \re^{- i \pi m /N} R_{1 \alpha} , \; \;
R_{2 \alpha} \rightarrow \re^{i \pi m /N} R_{2 \alpha} .
\label{tildeznfer}
\end{equation}
From these results and the definition (\ref{grep}), we deduce the transformation of the two original SU($N$)$_1$ 
WZNW fields. Under the ${\mathbb{Z}}_N$ symmetry, we have
 \begin{equation}
g_{1 } \rightarrow \re^{2 i  \pi m /N} g_{1}, \; \;
g_{2} \rightarrow \re^{-2 i \pi m /N} g_{2},
\label{znWZW}
\end{equation}
whereas $g_{1,2}$ are invariant under the ${\tilde {\mathbb Z}}_N$ symmetry.

Now we are in a position to reproduce the results of Ref. \onlinecite{griffin} which was based on 
CFT consistencies. First of all, since $g_{1,2 }$ transform in the fundamental representation of SU($N$), 
they should be related to the SU($N$)$_2$ WZNW primary field $G$ which transforms  in the same 
representation and  has scaling dimension $(N^2 - 1)/N(N+2)$ (see Appendix A).
Since the scaling dimension of $g_{1,2 }$ is $1 -1/N$, we need an additional operator 
in the ${\mathbb Z}_N$ CFT with scaling dimension $(N-1)/N(N+2)$, i.e., 
$\sigma_1$, $\sigma^{\dagger}_1$ or the disorder fields $\mu_1$, $\mu^{\dagger}_1$.
One way to eliminate the ambiguity is to use the transformation of the different fields under
the ${\mathbb{Z}}_N \times {\tilde {\mathbb{Z}}}_N$ symmetry.
This suggests the following identification:
\begin{eqnarray}
(g_1)_{\alpha \beta} \sim  G_{\alpha \beta} \sigma_1
\nonumber \\
(g_2)_{\alpha \beta} \sim  G_{\alpha \beta} \sigma^{\dagger}_1 .
\label{ident}
\end{eqnarray}
Indeed, the disorder operators $\mu_1$, $\mu^{\dagger}_1$ cannot appear in the decomposition
since $g_{1,2}$ are singlets under the ${\tilde {\mathbb Z}}_N$ symmetry. The occurence 
of $\sigma_1$ and $\sigma^{\dagger}_1$ in Eq. (\ref{ident}) are consistent with the transformation law
of  $g_{1,2}$ under the ${\mathbb Z}_N$ symmetry (\ref{znWZW}).
We note that the results (\ref{ident})  do not hold for $N=2$ which is a  special case because the fundamental representation of SU(2) is self-conjugate. In that case, the expression can be obtained using the fact that SU(2)$_2$ CFT is related to three decoupled  2D Ising models (see for instance Ref. \onlinecite{bookboso}).  
One important consequence of the identity (\ref{ident}) is that the one-step translation symmetry (\ref{trans}) becomes now:
 \begin{equation}
T_{a_0}: \; \; G \rightarrow \re^{2 i \pi /N} G  .
\label{transbis}
\end{equation}

Finally, the identification (\ref{ident}) can be generalized for 
all SU($N$)$_1$ primary fields $\varphi_l$ which transform in the antisymmetric representation of SU($N$) described by  a Young tableau with a single column and $l$ boxes ($ 1 \le l \le N-1$). \cite{griffin}
For the first SU($N$)$_1$ theory, these primary fields, i.e. $\varphi_{1l}$, are obtained by $l$ fusion
of $g_1$ by itself.  Using the result  (\ref{ident})  and the fusion rules of the ${\mathbb{Z}}_N$ parafermionic CFT,
one can derive the correspondence between $\varphi_{1,2 l}$ and 
 SU($N$)$_2$ and ${\mathbb{Z}}_N$ primaries. 
If we denote $\Phi_l$ the SU($N$)$_2$ primary field with scaling dimension 
$l(N+1) (N - l)/N(N+2)$ (see Appendix A) which transforms in the $l$th antisymmetric representation of SU($N$),
we find:
\begin{eqnarray}
\varphi_{1l} &\sim&  \Phi_l \sigma_l
\nonumber \\
\varphi_{2l} &\sim&  \Phi_l \sigma^{\dagger}_l ,
\label{identgen}
\end{eqnarray}
which is, of course, fine at the level of the scaling dimension since 
$ l(N - l)/N =  l(N+1) (N - l)/N(N+2) + l(N-l)/N(N+2)$.

We are now ready to find to express the $\lambda_2$-perturbation of Eq. (\ref{cft}) in the new basis. 
Since $\sigma_1 \sigma_1 \sim \sigma_2$, we obtain:
\begin{equation}
{\rm Tr}  \;  g_1 {\rm Tr}  \;  g^{+}_2  \sim  : {\rm Tr}   \;  G  \; {\rm Tr}   \; G^{+} : \sigma_2  \sim {\rm Tr}  
(\Phi_{\rm adj}) \sigma_2,
\label{V2rep}
\end{equation}
where $\Phi_{\rm adj}$ is the SU($N$)$_2$ primary field with scaling
dimension $2N/(N+2)$, which transforms in the adjoint representation of SU($N$). 
 In this derivation, we have used the definition of the adjoint primary field:\cite{knizhnik}
\begin{equation}
{\rm Tr}  (\Phi_{\rm adj}) = {\rm Tr} ( G^{+} T^{A} G T^{A} )= \frac{1}{2} \left( {\rm Tr}  \; G \;  {\rm Tr}   \;  G^{+}  - \frac{1}{N} {\rm Tr}  (G G^{+}) \right),
\label{TrPhiadj}
\end{equation}
and the identity
\begin{equation}
 \sum_A T^A_{\alpha \beta} T^A_{\gamma \rho} = 
\frac{1}{2} \left(\delta_{\alpha \rho} \delta_{\beta \gamma}
- \frac{1}{N}\; \delta_{\alpha \beta} \delta_{\gamma \rho} \right) .
\label{SUNident}
\end{equation}
As the result of Eq. (\ref{V2rep}) we obtain the following expression for $\lambda_2$-perturbation, $V_2$, of 
model (\ref{cft}): 
\begin{eqnarray}
V_2 \simeq \lambda_2 \int dx   \; {\rm Tr} (\Phi_{\rm adj}) \left( \sigma_2 + \sigma^{\dagger}_2 \right) .
\label{V2repfin}
\end{eqnarray}

In summary, the low-energy effective field theory (\ref{cft}) of weakly coupled SU($N$) Heisenberg chains 
can be reformulated in the basis (\ref{embedding}) with Hamiltonian: 
\bea
{\cal H} = W[SU(N)_2; G] + {\cal H}_{{\mathbb Z}_N}    - \tilde\lambda\int d x \;  \left( \Psi_{1L}  \Psi_{1R}  + H.c. \right)  
+ \lambda_2 \int dx   \; {\rm Tr} (\Phi_{\rm adj}) \left( \sigma_2 + \sigma^{\dagger}_2 \right) .
\label{cftnewbasis}
\eea
The crucial difference between $N=2$ and $N>2$ cases is that for $N=2$ there is no $\s_2,\s_2^{\dagger}$ operators and therefore two sectors of the theory decouple from each other. This does not happen for $N>2$ and this  determines the difference in physics. Recall now that for $N=2n$ the spectrum of  the original model (\ref{cft}) is invariant under a change of sign of the both coupling constants. Such sign change should be compensated by field transformations in Eq. (\ref{cftnewbasis}), but we managed to find them only for $N=4p$.

\section{Phases of the generalized two-leg SU($2n$) spin ladder}

In this section, we will investigate the IR physics of model (\ref{cftnewbasis})
to deduce the nature of the possible phases of  generalized two-leg SU($2n$) spin ladder.

\subsection{Field theory strong coupling approach}

To shed light on the possible phases, it might be interesting to first perform a strong-coupling approach directly to the continuum model (\ref{cft}) when both coupling constants $\lambda_{1,2}$ are of the order of the ultraviolet cut-off
and $|\lambda_1| >> |\lambda_2|$.
In this respect, let us consider the euclidean action corresponding to Eq. (\ref{cft}): \cite{knizhnik,dms}
\bea
{\cal S} = {\cal S}[SU(N)_1; g_1] +   {\cal S}[SU(N)_1; g_2]
+  \lambda_1  \int d^2x \; \Big[\mbox{Tr}(g_1g_2^+) + H.c.\Big] + \lambda_2
   \int d^2x  \;  \Big[\mbox{Tr}g_1\mbox{Tr}g_2^++ H.c.\Big],
\label{cftaction}
\eea
where the action of the SU($N$)$_k$ WZNW model is given by:
\bea
{\cal S}[SU(N)_k; g] &=& \frac{k}{8\pi} \int d^2 x \; {\rm Tr} \; (\partial^{\mu} g^{+} \partial_{\mu} g) 
+ \Gamma(g) \nonumber \\
\Gamma(g) &=& \frac{k }{12\pi}   \int_B d^3 y \; \epsilon^{\alpha \beta \gamma} 
 {\rm Tr} \; (g^{+} \partial_{\alpha} g g^{+} \partial_{\beta} g g^{+} \partial_{\gamma} g),
\label{WZW}
\eea
$g$ being an SU($N$) matrix field and $\Gamma(g)$ the WZNW topological term.

The results of the strong-coupling approach depend on the sign of $\lambda_1$ and we assume $N=2n$.

\subsubsection{$\lambda_1 <0$} 

Then the minimization of the $\lambda_1$ term in action (\ref{cftaction}) gives $g_1 = g_2 = G$ 
and the WZNW topological term (\ref{WZW}) is doubled.
The resulting effective action is therefore
\be
{\cal S}_{\rm eff} =  {\cal S}[SU(N)_2; G] + 2 \lambda_2  \int d^2 x \;  |\mbox{Tr}  \; G|^2.
\ee
Now, if $\lambda_2 <0$ then $ |\mbox{Tr}  \; G |$ should be maximal. For an SU($N$) matrix it leads to 
the conclusion that $G$ belongs to the center of SU($N$), i.e. ${\mathbb{Z}}_N$, with
$G =  \re^{2 i \pi k /N} I$, $k= 0, 1, \ldots, N -1$ and $I$ is the $N \times N$ identity matrix.
The one-step translation $T_{a_0}$ is spontaneously broken and 
the system has a finite order parameter with wave vector $2k_F = \pi/n$. 
If $\lambda_2 >0$, as this would be the case for the two-spin interchain exchange ladder where $\lambda_2 = - \lambda_1/N$, we get the condition $ {\rm Tr}  \;  G =0$. Here, as we will see, the order parameter has $4k_F$ wave vector.

\subsubsection{$\lambda_1 >0$}

A similar approach leads to $g_1 = - g_2 = G$ which is still an SU($N$) matrix if $N$ is even.
The resulting model becomes then 
\be
{\cal S}_{\rm eff} =  {\cal S}[SU(N)_2; G] - 2 \lambda_2  \int d^2 x \;  |\mbox{Tr} \; G|^2.
\ee
For $\lambda_2 >0$ we have again $G = \re^{2 i \pi k /N} I$ which corresponds 
to $2k_F$ order which is now staggered between the chains
since $g_1 = - g_2$.  For $\lambda_2 <0$ we get the condition ${\rm Tr} \; G = 0$ corresponding to an 4$k_F$ ordering.

\subsection{Integrable point $\lambda_2=0$. }

A weak-coupling approach can be performed by exploiting the fact that when $\lambda_2 =0$ model (\ref{cft}) becomes integrable and is related to the Fateev model  (\ref{intpert}). According to  Ref. \onlinecite{fateev}, 
the integrable deformation of the ${\mathbb{Z}}_N$ parafermions (\ref{intpert}) is a massive 
field theory with a mass gap $\Delta$ for any sign of its coupling constant ${\tilde \lambda}$ when $N$ is even.
The action explicitly breaks the ${\mathbb{Z}}_N$ symmetry and the exact spectrum consists of massive kink
excitations that result from degenerate ground states labelled by an odd integer $s=1,3, \ldots, N+1$.\cite{fateev,FZ}
One can then average over high-energy degrees of freedom represented 
by the theory (\ref{intpert}) and obtain an effective field 
theory for the SU($N$)$_2$ sector in the low-energy limit $E \ll \Delta$. This theory  describes magnetic 
excitations carrying quantum numbers of the SU($N$) group. Using the result (\ref{cftnewbasis}) 
of the previous section, we obtain the Hamiltonian density of this effective field theory: 
\begin{equation}
{\cal H}_{\rm eff} = \frac{2\pi v}{N+2}\Big(:I_R^A I_R^A: + :I_L^A I^A_L:\Big) + \eta_s  {\rm Tr}   \;  \Phi_{adj} , 
\label{eff}
\end{equation}
where 
$I_{R,L}^A$ are chiral  SU($N$)$_2$  currents and the mass $M$ of the ${\mathbb{Z}}_N$ particle 
serves as the upper cut-off. The coupling constant of the adjoint primary field $ \Phi_{adj} $ is
\begin{equation}
\eta_s = \lambda_2 \langle \sigma_2 +\sigma^{\dagger}_2 \rangle_s M^{d_{\Phi}},
\label{couplingeffmod}
\end{equation}
$d_{\Phi} = 2N/(N+2)$ being the scaling dimension of the adjoint primary field (see Appendix A).
When ${\tilde \lambda} > 0$, the vacuum average of the spin fields $\sigma_j$ in the ground state $s$ is known from  Ref. \onlinecite{bas}:  
\begin{eqnarray}
&& \langle \sigma_j \rangle_s  = \langle 0_s|\sigma_j|0_s \rangle = \frac{\sin\Big[\frac{\pi(j+1)s}{N+2}\Big]}{\sin\Big(\frac{\pi s}{N+2}\Big)}(M/4)^{2h_j} e^{Q_j}\label{vev}\\
&& Q_j = \int_0^{\infty}\frac{d t}{t}\Big\{\frac{\sinh(tj)\sinh[(N-j)t]}{\sinh(Nt)\sinh[(N+2)t]}- 2h_j e^{-2t}\Big\},\nonumber
\end{eqnarray}
where $h_j = j(N-j)/N(N+2)$. In particular, from Eq. (\ref{vev}) we see that $\la \s_1\ra_s = \la\s_{N-1}\ra_s \equiv \la\s_1^{\dagger}\ra_s$. According to Eq. (\ref{ident}), this means that for $\lambda_1 < 0$, i.e., $\tilde\lambda \sim -\lambda_1 >0$,
we have $g_1 = g_2$, as we envisaged from the strong-coupling approach of the previous subsection.

Let us now return to the problem of finite $\lambda_2$. Our task  is to analyse the infrared physics of the SU($N$)$_2$ model (\ref{eff}) perturbed by its adjoint primary field.
Since it is a strongly relevant perturbation, we may  expect that the magnetic SU($N$) sector is always gapped. 

 Fine details of the spectrum, however, depend on the sign and magnitude of coupling constant $\eta_s$ (\ref{couplingeffmod}). According to the result (\ref{vev}) these depend on the ground state of model 
 (\ref{intpert}) $|0\ra_s$. We suggest that $s$  is selected  in such a way   that the ground-state energy of (\ref{eff})  is the lowest. It is reasonable to think that this corresponds to largest spectral gaps.   

 To determine a qualitative dependence of  the spectrum  on $\eta_s$ 
it is useful to consider a direct semiclassical approach to the interacting Hamiltonian 
density (\ref{eff}) using the identity (\ref{TrPhiadj}):
\begin{equation}
{\cal H}_{\rm int} =  \eta_s {\rm Tr}  \; G  \;  {\rm Tr}   \;  G^{+},
\label{adjointpertsem}
\end{equation}
where $G$ is now an SU($N$) matrix. When $ \eta_s <0$, the minimization selects the center group of SU($N$):
\begin{equation}
G =    \exp(\ri  2 \pi k/N) I       , \; \,  \eta_s < 0 ,
\label{min1adj}
\end{equation}
where $k = 0, \ldots, N-1$ and the solution breaks spontaneously the one-step translation symmetry
$T_{a_0}$ (\ref{transbis}). The ground state is thus $N$-fold degenerate.

When $ \eta_s >0$, the minimization gives then the condition that G is an SU($N$) matrix with
the constraint: ${\rm Tr}  \;  G =0$. The general solution for $N=2n$ takes then a Grassmanian form: \cite{affleck} 
\begin{equation}
G =  \exp(\ri  2 \pi k/N) \; U^+ {\rm diag}(1,...1,-1,...-1) U, \; \,  \eta_s > 0 ,
 \label{grassmanmin2}
\end{equation}
$U$ being a general unitary U($N$) matrix. 
As shown in Ref. \onlinecite{affleck}, within this semiclassical approach, model (\ref{eff}) with even $N$ 
becomes  equivalent to the Grassmanian sigma model on $U(N)/[U(N/2)\times U(N/2)]$ manifold
with a trivial topological term $\theta = 2 \pi$. This model describes  
a fully gapped phase.

 Although for both signs of $\eta_s$ we obtain gapped spectra, the gaps for the Grassmanian sigma model are expected to be smaller. This becomes clear at large $N>>1$ since $1/N$ serves as a coupling constant for the sigma model and hence the gaps are exponentially small in $N$: $m \sim \exp(- \mbox{const}N)$. At the same time for $ \eta_s <0$ the gaps are algebraic in $1/N$.

From these results, obtained within the semiclassical analysis, we can now determine a phase diagram  of the 
generalized two-leg SU($2n$) spin ladder.

\subsection{$\lambda_1 <0$}

We first assume that $\lambda_1 <0$ so  that $\tilde\lambda \sim -\lambda_1 >0$ and we can use the 
result (\ref{vev}) for the vacuum expectation
values of the  $\sigma_j$ fields. As we have mentioned above, the coupling constant 
$\eta_s$ (\ref{couplingeffmod}) of the low-energy theory (\ref{eff}) depends on the vacuum of model (\ref{intpert}). The degeneracy is  removed by a selection of $\eta_s$ which yields the lowest energy ground state energy for 
model (\ref{eff}).

To establish a relationship between $s$ and $\eta_s$ we recall that the vacuum expectation values (\ref{vev}) enjoy the property:
$\langle \sigma_j\rangle_s =  (-1)^{1+s} \langle \sigma^{\dagger}_j \rangle_s = \langle \sigma^{\dagger}_j \rangle_s$ since
$s$ is odd. Then substituting Eq. (\ref{vev}) into Eq. (\ref{couplingeffmod}) we get for $N=2n$:
\be
\eta_{s=2k+1} \sim \lambda_2\Big(1+2\cos\Big[\frac{\pi(2k+1)}{n+1}\Big]\Big), ~~ k= 0,1,...n. \label{const}
\ee

\subsubsection{$\lambda_2 <0$}

When $\lambda_2 <0$, the result (\ref{const}) suggests that the true ground state of model (\ref{eff}) corresponds to $s=1$.
Indeed,  (i)  this state has the largest value of the coupling $\eta_s$, (ii) $\eta_{s=1} < 0$ which according to the semiclassical arguments presented above,  corresponds to a larger gap than for a positive coupling. The semiclassical analysis conducted above predicts  a spin gap phase with $N$-fold ground-state degeneracy which stems from the complete breaking the one-step translation symmetry $T_{a_0}$.  Using the identification (\ref{ident}), we deduce the result:
\begin{equation}
{\rm Tr} \; g_1 \pm {\rm Tr} \;  g_2 \sim  {\rm Tr} \;  G  \;  (\sigma_1 \pm \sigma^{\dagger}_1 ),
\label{dimerop}
\end{equation}
and since $\langle \sigma_1 \rangle_{s=1} =  \langle \sigma^{\dagger}_1 \rangle_{s=1} \ne 0$ from Eq. (\ref{vev}), we have:
$\langle  {\rm Tr} \; g_1 \rangle = \langle  {\rm Tr} \; g_2 \rangle  \ne 0$.
A lattice order parameter can then be deduced by means of the result of Appendix B (see Eq. (\ref{latticebondTrprimary}) 
with $m=1$):
\begin{eqnarray}
\langle {\cal O}^{{\rm u} \dagger}_{2k_F} (n)\rangle = \re^{- i \frac{2 \pi n}{N}} 
\langle \sum_{l=1}^{2} S^{A}_{l,n}  S^{A}_{l, n+1} \rangle
\sim \langle  {\rm Tr} \; g_1  + {\rm Tr} \;  g_2  \rangle  \ne 0 .
\label{latticebondphase1}
\end{eqnarray}
Thus we have a uniform $2k_F$ VBS phase which is in phase between the legs of the two-leg ladder
and the wave vector is thus $(\pi/n, 0)$. This $N=2n$-fold degenerate phase breaks spontaneously
the $T_{a_0}$ symmetry as expected from the semiclassical approach.

\subsubsection{$\lambda_2 > 0$}

In this case according to Eq. (\ref{const}) for even $n$  the most negative coupling constant corresponds to a single vacuum $k = n/2$. For odd $n$ there are two degenerate vacua $k = (n\pm 1)/2$. 

Although the spectrum of model (\ref{eff}) remains the same as for $\lambda_2 <0$, the order parameters change. Indeed,   from Eq. (\ref{vev}) we have 
\be
\la \s_1\ra_{2k+1} \sim \cos\Big(\frac{\pi(2k+1)}{2n+2}\Big),
\ee
and for $k=n/2$ this vacuum expectation value vanishes. From the correspondence (\ref{dimerop}), we deduce that
the $2k_F$ order parameter vanishes as well:   $\langle  {\rm Tr} \; g_1  + {\rm Tr} \;  g_2  \rangle = 0$, in agreement
with the strong-coupling approach. However, since $G^2 =  \exp(\ri  4 \pi k/N) I$ from Eq. (\ref{min1adj}) and $ \langle \sigma_2 \rangle_{s=n+1} \ne 0$, for $N=4p$ 
we have instead the formation of a $4k_F$ VBS phase:
\begin{eqnarray}
\langle {\cal O}^{\dagger}_{4k_F} (n)\rangle = \re^{- i \frac{4 \pi n}{N}} \langle \sum_{l=1}^{2} S^{A}_{l,n}  S^{A}_{l, n+1} \rangle
\sim \langle  {\rm Tr} \;  \varphi_{11} + {\rm Tr} \; \varphi_{21} \ra  
\sim \langle  {\rm Tr} \;  G^2  \ra  \ne 0 ,
\label{latticebondphase2}
\end{eqnarray}
where we have exploited the result (\ref{latticebondTrprimary}) with $m=2$ of Appendix B and Eq. (\ref{identgen}). 
The $4k_F$ VBS phase has wave vector $(2\pi/n, 0)$ and a $n$-fold ground-state degeneracy.

For $n$ odd we have 
\bea
\la \s_1\ra_{n,n+2} \sim \pm \sin\Big(\frac{\pi}{2n+2}\Big),
\eea
and $\la\s_1\ra$ does not vanish for odd $n$. In this respect, one might expect the existence of a $2k_F$ VBS phase
described by the order parameter (\ref{latticebondphase1}).
In contrast, the strong-coupling approach of Sec. IV A leads to the formation of a $4k_F$ VBS phase as in the
$n$ even case. One cannot exclude the occurence of a quantum phase transition between weak and
strong coupling regimes when $n$ is odd. However, the degeneracy between the $s=n$ and $s=n+2$ states 
might not be protected and with a result that the true ground state is a symmetric combination of the two states with opposite signs of $\s_1$ so that the resulting average vanishes: $\la \s_1\ra = 0$.
Such scenario would support the  $4k_F$ VBS phase as in the $n$ even case. 
Numerical calculations on the two-leg SU(6) spin ladder are clearly called for to shed light on the issue
of a quantum phase transition.

\subsection{$\lambda_1 > 0$}

We now turn to  $\lambda_1 > 0$ case ($\tilde\lambda<0$). The ${\mathbb{Z}}_N$ model  (\ref{intpert}) is still a massive
field theory. Unfortunately,  the vacuum expectation values of the order parameters (\ref{vev}) for $\tilde\lambda <0$ are not known. However, when $N=4p$ (even $n$), one can perform the transformation $\Psi_{1 L,R} \rightarrow \ri \Psi_{1 L,R}$ 
to change the sign of $\tilde\lambda$ in model  (\ref{intpert}). 
From the fusion rules of the  ${\mathbb{Z}}_N$ parafermionic theory \cite{para}
we have $\sigma_1 \mu_1 \sim \Psi_{1 L}$ and $\sigma_1 \mu^{\dagger}_1 \sim \Psi_{1 R}$, we deduce that
\be
\sigma_1 \rightarrow \ri \sigma_1, ~~ \sigma_2 \rightarrow - \sigma_2,  \label{trans1}
\ee
and thus $\la \s_1\ra_{s} = - \la \s^{\dagger}_1\ra_{s}$. As shown in Appendix C, the ${\mathbb{Z}}_4$ parafermions 
CFT can be  described by a bosonic field theory with central charge $c=1$. 
In particular, one can establish the mapping $\sigma_2 \rightarrow - \sigma_2$  in the $N=4$ case directly by means of this bosonized description.

The transformation (\ref{trans1}) maps the spectrum of model (\ref{cft}) with couplings $\lambda_1 >0, \lambda_2$ onto the spectrum with couplings $-\lambda_1, - \lambda_2$. The former one in Eq. (\ref{trans1}) means that the order parameter in the region $\lambda_1>0,\lambda_2>0$ is $\la\mbox{Tr}(g_1 -g_2)\ra$  using the correspondence (\ref{dimerop}).
It leads to the formation of a staggered $2k_F$ VBS phase with a $N$-fold degeneracy:
\begin{eqnarray}
\langle {\cal O}^{{\rm stag} \dagger}_{2k_F} (n)\rangle = \re^{- \frac{2 i \pi n}{N}} \langle \sum_{l=1}^{2} (-1)^{l+1} 
S^{A}_{l,n}  S^{A}_{l, n+1} \rangle
\sim \langle  {\rm Tr} g_1  -  {\rm Tr} g_2  \rangle  \ne 0 .
\label{latticebondphase3}
\end{eqnarray} 
The emergence of a staggered phase is consistent with the strong-coupling approach in the $\lambda_1 > 0, \lambda_2 > 0$ 
region.

\begin{figure}[!ht]
\centering
\includegraphics[width=0.5\columnwidth,clip]{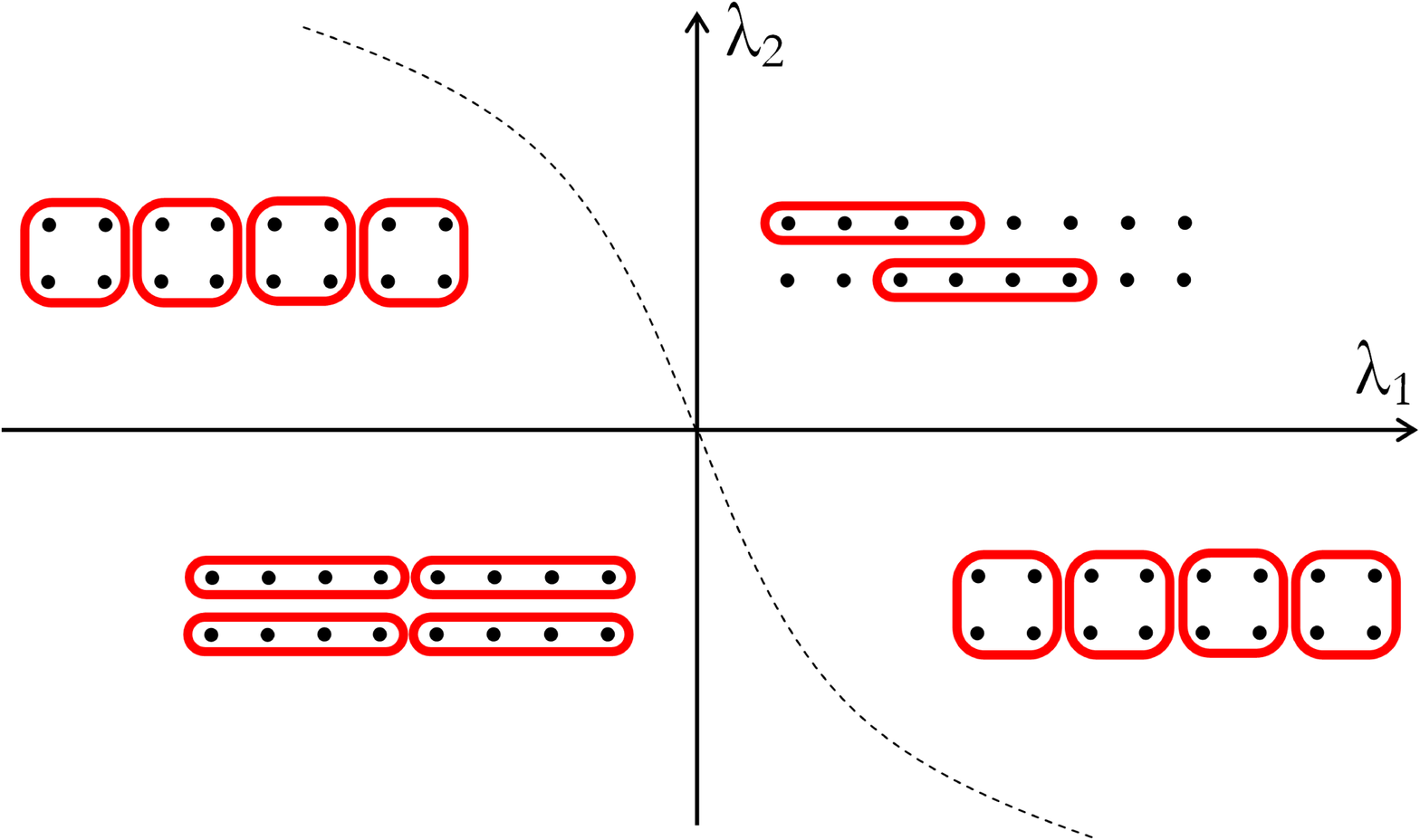}
\caption{Phase diagram of model (\ref{cft}) for $N=4$ by means of a field-theory analysis when
$|\lambda_1| >> |\lambda_2|$ and a mean-field approach for 
$|\lambda_2| >> |\lambda_1|$. A similar phase diagram is obtained in the even $n$ case where the 
plaquette phase is a $n$-site rectangular singlet phase.}
\label{fig:phasdiag}
\end{figure}

In contrast, the region $\lambda_1 >0, \lambda_2 <0$ is occupied by $4k_F$ VBS with a $N/2$-fold degeneracy as in Eq. (\ref{latticebondphase2}).   For $N=4$, it corresponds to a plaquette phase which is known to appear in the standard two-leg SU(4) spin ladder. \cite{mila,wu}
A summary of the results obtained for even $n$ can be found in Fig. 1.

When $N =4p+2$, the vacuum expectation values (\ref{vev}) of the Fateev model (\ref{intpert}) with $\tilde\lambda <0$ 
are not known so we cannot conclude as before unfortunately. However, from Eq. (\ref{cft}), we saw  that
the change of sign $\lambda_{1,2} \rightarrow - \lambda_{1,2}$ can be absorbed by the redefinition $g_1 \rightarrow - g_1$. The spectrum of model (\ref{cft})  for $\lambda_1 >0, \lambda_2$ is thus
related to that of with couplings $\lambda_1<0, - \lambda_2$. 
We thus expect the formation of a staggered $2k_F$ (respectively $4k_F$) VBS phase 
when $\lambda_2 > 0$ (respectively $\lambda_2 <0$) as in the even $n$ case.

\subsection{Mean-field analysis in the $|\lambda_2| >> |\lambda_1|$ regime}

So far, we have exploited the existence of an integrable line  $\lambda_2 = 0$ in model (\ref{cft}) to find the possible
phases of generalized two-leg SU($2n$)  spin ladders. As emphasized before,
such approach is valid in the  $|\lambda_1| >> |\lambda_2|$ regime.
When $\lambda_1 = 0$, the strongly relevant perturbation  in Eq. (\ref{cftnewbasis})  is not an integrable and 
is thus difficult to analyse.  In the regime $|\lambda_2| >> |\lambda_1|$, we expect the formation of spectral gaps in the
problem. One can then investigate the resulting IR physics in this regime by means of a simple mean-field decoupling approach:
${\cal H}_{\rm mf} = {\cal H}_1 + {\cal H}_2$ with:
\begin{eqnarray}
{\cal H}_1 &=& {\cal H}_{{\mathbb Z}_N}   +
\lambda_2 \int dx   \; \langle {\rm Tr} (\Phi_{\rm adj}) \rangle \left( \sigma_2 + \sigma^{\dagger}_2 \right) 
\nonumber \\
{\cal H}_2 &=&
\frac{2\pi v}{N+2}\Big(:I_R^A I_R^A: + :I_L^A I^A_L:\Big) 
+ \lambda_2 \int dx \; \langle \sigma_2 + \sigma^{\dagger}_2 \rangle \; {\rm Tr} (\Phi_{\rm adj})  .
\label{meanfield}
\end{eqnarray}
The perturbation in the ${\mathbb{Z}}_N$ sector with the $\sigma_2$ spin field with scaling dimension $2(N-2)/N(N+2)$
is more relevant than the one in the Fateev model (\ref{intpert}). In the regime $|\lambda_2| >> |\lambda_1|$, one
can thus ignore the latter perturbation. The field theory ${\cal H}_1$ is not integrable except in the $N=4$ case
where it is equivalent to the sine-Gordon model with $\beta^2 = 2 \pi/3$ (see Appendix C).
The $\sigma_2$ spin perturbation gives a mass gap to the  ${\mathbb{Z}}_N$ degrees of freedom and, 
when $\lambda_2 <0$, we have $ \langle \sigma_2 + \sigma^{\dagger}_2 \rangle > 0$ since $ \langle {\rm Tr} (\Phi_{\rm adj}) \rangle > 0$.
The perturbation in the SU($N$) sector, described by ${\cal H}_2$, 
becomes equivalent to model (\ref{eff}) with $\eta_s <0$. The one-step translation symmetry is then spontaneously
broken and a 2k$_F$ VBS phase emerges in the regime $ |\lambda_2| >> |\lambda_1|$.
When $\lambda_1 >0$ for instance, there is thus a competition between 4k$_F$  and 2k$_F$  VBS phases 
which is marked by the transition line in Fig. 1. The full determination of the transition lines is beyond the scope
of our approach.

A strong-coupling analysis of the lattice standard  antiferromagnetic two-leg 
SU($2n$) ladder is presented in Appendix D. In Fig. 1, it corresponds to the line $\lambda_2 = - \lambda_1/N$ with
$\lambda_1 \rightarrow \infty$. For $N=4$ and $N=6$, we find respectively plaquette and trimerized phases, i.e., $4k_F$ VBS phases. The lattice strong-coupling analysis confirms thus the 
prediction of our field-theory approach presented in Sec. IV  when $|\lambda_1| >> |\lambda_2|$.
We thus conclude that  the $4k_F$ VBS phase occurs in the phase diagram of the two-leg SU(4) and SU(6) spin ladders 
for a sufficiently strong $J_{\perp} >0$. Whether this phase is stabilized upon switching on a small 
interchain coupling or remplaced by a competing $2k_F$ VBS phase  is an interesting question which 
cannot be answered unfortunately within our approach. In this respect, large-scale numerical simulations are necessary to
shed light on this intriguing possibility.
When $n >4$,  an SU($2n$)$_1$ quantum criticality is expected from 
the lattice strong-coupling approach of Appendix D. A quantum phase transition should occur between weak and strong coupling regimes of the standard SU($2n$) spin ladder with  $J_{\perp} >0$.

\section{Conclusions}

 We have studied the zero-temperature phase diagram of the SU($2n$) Sutherland antiferromagnetic ladder with weak but generic interchain interactions. We have found that for a translationally invariant ladder with inversion symmetry this phase diagram is controlled by only two parameters $\lambda_{1,2}$ in the continuum limit. These are coupling constants of the two most relevant operators allowed by the symmetries. 
Our field-theory analysis exploits the existence of an integrable massive field theory of ${\mathbb{Z}}_N$ parafermions
along the $\lambda_2 =0$ line. In the vicinity of that line, i.e. $|\lambda_1| >> |\lambda_2|$, our approach
leads to the conclusion that the phase diagram contains only VBS phases characterized by local order parameters. In this respect, there are no topological phases as for $n=1$,i.e., a two-leg SU(2) spin ladder.  There are two types of VBS phase with wave vectors $(\pi/n, 0)$ or $(\pi/n,\pi)$ ($2k_F$ VBS) and $(2\pi/n, 0)$  ($4k_F$ VBS). 
The $4k_F$ VBS can be viewed as a cluster of $2n$ spins put in an SU($2n$) singlet with a $n$-fold ground-state degeneracy. For $N=4$, it corresponds to a plaquette phase which is known to appear in the standard two-leg SU(4) 
spin ladder. \cite{mila,wu}
Our results for $N=6$ suggest that a cluster phase of six spins, leading to trimerization, should occur in the phase
diagram of the two-leg SU(6) spin ladder. The latter case is directly relevant to the insulating phase 
of double tube of ytterbium $^{173}$Yb ultracold atoms which can be engineered by considering double-well optical lattices.
\cite{porto}
In this respect, it will be interesting to investigate numerically the latter system to confirm the existence of such a 
cluster phase for $N=6$ by means of numerical methods for SU($N$) magnet as in Ref. \onlinecite{nataf}. 
Finally, the field-theory analysis of two-leg spin ladder with SU($2n+1$) spins is different and will be presented
elsewhere.

\begin{acknowledgements}
The authors are grateful to S. Capponi, V. Gurarie, P. Baseilhac, V. Bois, V. Fateev, S. Manmana, F. Mila,
and P. Nataf for very useful discussions. 
AMT was funded by US DOE under contract number DE-AC02 -98 CH 10886.
\end{acknowledgements}

\appendix

\section{Quadratic Casimir of SU($N$)}

In this Appendix, we give some conformal data of some SU($N$)$_k$  primary fields which 
appear in the conformal embedding approach of Sec. III.
The scaling dimension of a SU($N$)$_k$ primary field which transforms in some representation $R$ 
of the SU($N$) group is given by: \cite{knizhnik}
\begin{equation}
\Delta_R = \frac{2 C_R}{N+k},
\label{dimensioprimary}
\end{equation}
where $C_R$ is the quadratic Casimir in the representation $R$.
Its expression can be obtained from the general formula where $R$ is written as a Young tableau:
\begin{equation}
C_R = T^aT^a = \frac{1}{2}[l(N-l/N) +\sum_{i=1}^{n_{row}} b_i^2 - \sum_{i=1}^{n_{col}}a_i^2] 
\label{Cas}
\end{equation}
for Young tableau of $l$ boxes consisting of $n_{row}$ rows of length $b_i$ each 
and $n_{col}$ columns of length $a_i$ each.
For instance, we get $C_R = (N^2 -1)/2N$ for the fundamental representation, 
$C_R = N$ for the adjoint representation, $C_R (k) = k(N+1) (N - k)/2N$ for the kth basic antisymmetric representation
made of a Young tableau with a single column and $k$ boxes, and $C_R = N - 2/N +1$ for the symmetric 
representation with dimension $N(N+1)/2$.

 \section{Lattice representation of SU($N$)$_1$ primary fields}
 
We investigate here the lattice representation of SU($N$)$_1$ primary fields which transform
in the basic antisymmetric representations  of SU($N$). The results are important for the identification 
of the phases of the generalized two-leg SU($2n$) spin ladder.
 
 Let us consider an SU($N$) spin chain in the fundamental representation of SU($N$). As recalled in the introduction,
the model is integrable and displays a quantum critical behavior in the SU($N$)$_1$ universality
 class with central charge $c= N-1$. \cite{sutherland,affleck}
At low-energy, the lattice SU($N$) operators in the continuum limit are
described by: \cite{affleck,assaraf}
\begin{equation}
S^{A}_{n} \simeq J^{A}_{L} +  J^{A}_{R} + \sum_{m=1}^{N-1} \mbox{e}^{ \ri 2m k_F x} N_m^{A},
\label{spinopappen}
\end{equation}
where $k_F = \pi/Na_0$, $x=n a_0$, and $J^{A}_{L,R}$ are the left and right SU($N$)$_1$ currents.
The $2m k_F$ parts of this decomposition are related to the $m=1, \ldots, N-1$ 
SU($N$)$_1$ primary field $\Phi_{m}$ with scaling dimension $m(N-m)/N$ which transforms
in the antisymmetric representation  of SU($N$)  made of a Young tableau with a single column and $m$ lines:
\begin{equation}
 N_m^{A} = \lambda_m  \; {\rm Tr} ( \Phi_{m} T_m^A),
\label{spinopprimaryappen}
\end{equation}
where $T_m^A$ are SU($N$) generators in the $m$th basic antisymetric representation of the SU($N$) group
and $\lambda_m$ non-universal real constants.
It is interesting to get some lattice interpretation of ${\rm Tr} \; \Phi_{m}$ in terms of the 
original lattice SU($N$) spin operators. In this respect, let us introduce lattice $2mk_F$ bond SU($N$) operators:
\begin{eqnarray}
{\cal O}^{\dagger}_{2mk_F} (n) = \re^{- \ri \frac{2 \pi m n}{N}} S^{A}_{n}  S^{A}_{n+1}   ,
\label{latticebondop}
\end{eqnarray}
with $m=1, \ldots, N-1$. The continuum description of these operators can be obtained from the identification (\ref{spinopappen}):
\begin{eqnarray}
{\cal O}^{+}_{2mk_F} (n) \simeq N_m^{A} (x)  \left[  J^{A}_{L} (x + a_0) +  J^{A}_{R} (x +a_0) \right] 
+  \re^{\ri \frac{2 \pi}{N}} \left[  J^{A}_{L} (x) +  J^{A}_{R} (x) \right]  N_m^{A} (x+a_0)  .
\label{latticebondopcont}
\end{eqnarray}

At this point, it is useful to recall the defining operator product expansions (OPE) for the SU($N$)$_1$  primary
fields:\cite{dms}
\begin{eqnarray}
J^{A}_{L} (z) \left(\Phi_{m}\right)_{r,s} (w, \bar w) &\sim& - \frac{1}{2 \pi(z-w)} \left(T_m^A\right)_{r,p} \left(\Phi_{m}\right)_{p,s} (w, \bar w) \nonumber \\
J^{A}_{R} (\bar z) \left(\Phi_{m}\right)_{r,s} (w, \bar w) &\sim& \frac{1}{2 \pi(\bar z - \bar w)} \left(T_m^A\right)_{p,s} \left(\Phi_{m}\right)_{r,p} (w, \bar w) .
\label{OPEsprimarySUNdef}
\end{eqnarray}
Using this result together with the definition (\ref{spinopprimaryappen}), we get the following OPE:

\begin{eqnarray}
\re^{\ri \frac{2 \pi}{N}}  \left[ J^{A}_{L} (z) + J^{A}_{R} (\bar z) \right] N_m^{A} (w, \bar w) 
+ N_m^{A} (z, \bar z)  \left[ J^{A}_{L} (w) + J^{A}_{R} (\bar w) \right] &\sim&  \nonumber \\
- \frac{\lambda_m C_m}{2 \pi} \left( e^{i \frac{2 \pi}{N}} - 1 \right)  \left[ \frac{1}{z-w} - \frac{1}{\bar z- \bar w} \right]
{\rm Tr} \;  \Phi_{m} (w, \bar w) ,
\label{OPEsprimarySUNcurr}
\end{eqnarray}
where $C_m$ is the quadratic Casimir of the $m$th antisymmetric representation of the SU($N$) group.
We then deduce the lattice representation of ${\rm Tr} \; \Phi_{m}$ in terms of the $2mk_F$ bond SU($N$) operators:
\begin{eqnarray}
{\cal O}^{\dagger}_{2mk_F} (n) = \re^{- \ri \frac{2 \pi m n}{N}} S^{A}_{n}  S^{A}_{n+1}  \sim {\rm Tr} \;  \Phi_{m} (x) .
\label{latticebondTrprimary}
\end{eqnarray}

\section{A special case of the SU(4) group}

For $N=4$ one can exploit the fact that SU(4) group is  isomorphic to O(6). The Kac-Moody algebras SU(4)$_k$
and O(6)$_k$ are equivalent and for $k=1$ and $k=2$ it is possible to employ Abelian bosonization and Majorana fermions
techniques.\cite{bookboso}
 
\subsection{${\mathbb{Z}}_4$ Fateev model and sine-Gordon model}
 
In particular, the ${\mathbb{Z}}_4$  parafermion CFT has central charge $c=1$ and admits 
a bosonized description in terms of a compactified boson field $\chi$ at radius $R=\sqrt{3/2\pi}$ 
($\chi \sim \chi + \sqrt{6\pi}$) which
lives on a ${\mathbb{Z}}_2$ orbifold: $\chi \sim - \chi$. \cite{yang}
Some primary fields of the ${\mathbb{Z}}_4$ CFT have a simple bosonic representation in terms
of a vertex operator as the $\sigma_2$ field and the first thermal operator $\epsilon_1$:
\begin{equation} 
\s_2 \sim \cos(\sqrt{2 \pi/3}\;  \chi), ~~ \epsilon_1 \sim \cos(\sqrt{8 \pi/ 3} \; \chi),
\label{freebosonZ4para}
\end{equation} 
which have respectively scaling dimension $1/6$ and $2/3$ as it should. 
The spin field operators $\s_1, \s_1^+$ with scaling dimension 
$1/8$ are related to the twist fields of the $c=1$ ${\mathbb{Z}}_2$ orbifold CFT.\cite{dms}

Using this bosonization approach, the Fateev model (\ref{intpert}) can be
shown to be equivalent to the $\beta^{2} = 6 \pi$ sine-Gordon theory with Hamiltonian density:\cite{fateev,FZ,gogolin}
\bea
{\cal H}_{{\mathbb{Z}}_4} = \frac{1}{2} \left( (\partial_x \chi)^2 + (\partial_x {\tilde \chi})^2 \right) - {\tilde \lambda} \cos(\sqrt{6\pi } \; \chi),
\label{SGZ4}
\eea
${\tilde \chi}$ being the dual field of $\chi$.

According to the exact solution of model  (\ref{intpert}),  there are three degenerate vacua 
where $\s_2$ acquires expectation values  (see Eq. (\ref{vev})):
\bea
\la 0|\s_2|0\ra_s \sim 1 +2\cos(\pi s/3), ~~ s=1,3,5,
\label{s2vevn4}
\eea
corresponding to $2,-1,2$. 
The sine-Gordon model (\ref{SGZ4}) at $\beta^{2} = 6 \pi$ leads to the pinning of the bosonic field
at the minima of the potential when ${\tilde \lambda} >0$ : $\langle \chi \rangle = p \sqrt{2\pi/3}$, $p$ being
integer. Taking into account the radius of the boson which leads to the identification $\chi \sim \chi + \sqrt{6\pi}$,
we see that  the sine-Gordon model (\ref{SGZ4}) has three degenerate ground states with:
$\langle \chi \rangle = 0,  \sqrt{2\pi/3}, 2 \sqrt{2\pi/3}$. Using the bosonic representation (\ref{freebosonZ4para}),
we find that the vacuum expectation values of the $\s_2$ operator give three values with two of them being equal
as the exact result (\ref{s2vevn4}).
Using this bosonic approach, we can also determine how result  (\ref{s2vevn4}) is modified when ${\tilde \lambda} < 0$.
The change of sign ${\tilde \lambda} \rightarrow - {\tilde \lambda}$ can be obtained by the mapping:
$\chi  \rightarrow \chi +   \sqrt{\pi/6} + q   \sqrt{2\pi/3}$, $q$ being integer. The value of $q$ is fixed by the requirement
that the first thermal operator should be invariant under the transformation since the ${\mathbb{Z}}_4$ symmetry
should not be broken when ${\tilde \lambda} < 0$. 
The bosonization result (\ref{freebosonZ4para}) gives $q=1$ and therefore:
\be
\chi  \rightarrow \chi + \sqrt{3\pi/2} .
\ee
We thus deduce that under the change of sign ${\tilde \lambda} \rightarrow - {\tilde \lambda}$, the $\s_2$ operator
transform as: $\s_2  \rightarrow -\s_2$. We recover the result (\ref{trans1}) from the bosonization
approach for $N=4$.

\subsection{Abelian bosonization approach in the $N=4$ case}

Using this bosonization description of ${\mathbb{Z}}_4$ parafermions,  one can investigate 
a free-field representation of model (\ref{cftnewbasis}) when $N=4$.
The central charge of the SU(4)$_2$ is $c =5$ and this model might be described in terms of five bosonic fields. 

First of all, the currents of the SU(4)$_1$ $\sim$ SO(6)$_1$  CFT, which has central charge $c=3$, can be expressed in terms of six Majorana fermions $\chi_a$:
\bea
j^{ab} = \ri\chi_a\chi_b, ~~ 1 \le a < b \le 6.
\eea
Then the SU(4)$_2$ currents $J^{ab}$, being the sum of two SU(4)$_1$  ones, can be bosonized since
two Majorana fermions can be expressed in terms of a single bosonic field:
\bea
&& J^{ab} = \ri\chi_a\chi_b + \ri\eta_a\eta_b = \frac{\ri \kappa_a\kappa_b}{\pi a_0} [\cos(\sqrt{4\pi}\varphi_a)\cos(\sqrt{4\pi}\varphi_b)+ \sin(\sqrt{4\pi}\varphi_a)\sin(\sqrt{4\pi}\varphi_b)] = \nonumber\\
&& \frac{\ri\kappa_a\kappa_b}{\pi a_0}\cos[\sqrt{4\pi}(\varphi_a - \varphi_b)] ,
\label{currents}
\eea
where $\kappa_a$ are Klein factors. Since the center of mass of these bosonic fields drops out from this expression, we get the central charge 5. Thus we have a proof that SU(4)$_2$ can be bosonized. 
We suggest  the following bosonic form of model (\ref{eff}):
\bea
{\cal L}_{eff} = \frac{1}{2}\sum_{a=1}^6(\p_{\mu}\Phi_a)^2 - \gamma\sum_{a>b}\cos\Big[\sqrt{\frac{8\pi}{3}}(\Phi_a - \Phi_b)\Big].
\label{model2}
\eea
The center of mass field 
\bea
\Phi_0 = \frac{1}{\sqrt 6}\sum_a\Phi_a
\label{centermass}
\eea
is decoupled from the interaction and is redundant. Here, however,  it is important to obsverve that 
the fields cannot be the same as in  Eq. (\ref{currents}) because otherwise Tr$\Phi_{adj}$ will not be local with respect to $J$. 
Using Eq. (\ref{freebosonZ4para}) and the identification (\ref{SGZ4}),  we deduce a bosonic description
of model (\ref{cftnewbasis}) when $N=4$ with Lagrangian density:
\bea
{\cal L} = \frac{1}{2}(\p_{\mu}\chi)^2 + \frac{1}{2}\sum_{a=1}^6(\p_{\mu}\Phi_a)^2 - 
\lambda_1\cos(\sqrt{6 \pi} \chi) - 
\lambda_2\cos(\sqrt{2 \pi/3} \chi)
\sum_{a>b}\cos\Big[\sqrt{\frac{8\pi}{3}}(\Phi_a - \Phi_b)\Big]. 
\label{full}
\eea

From Eq. (\ref{s2vevn4}), we see that the operator 
$\cos(\sqrt{2 \pi/3} \chi)$ has three vacuum expectation values (two of which are equal to each other). 
The minimal ground-state energy is achieved when the sign of $\gamma = \lambda_2\la\cos(\sqrt{2 \pi/3}\chi )\ra$ is positive.
At $|\lambda_1| >> |\lambda_2|$ we average over $\chi$ field first and obtain the effective theory (\ref{model2}) with $\gamma >0$.  The latter can be studied either by $1/N$-expansion when we artificially extend the summation over 
isotopic indices to $N$.
Then we decouple the interaction via Hubbard-Stratonovich transformation and consider a saddle point:
\bea
{\cal L}_{\rm eff} = \frac{|\Delta|^2}{2\gamma} + \sum_a \Big[\Big(\Delta \re^{\ri\sqrt{8\pi/3}\Phi_a} + H.c.\Big) + \frac{1}{2}(\p_{\mu}\Phi_a)^2\Big],
\eea
($\gamma \sim \lambda_2$). 
The saddle point represents $N$-copies of the $\beta^2 = 8\pi/3$ sine-Gordon theory; its spectrum consists of kink, antikink and one breather with mass $\sqrt 3$ of the kink's mass.

 \subsubsection{Confinement of the heavy particles}
 
The structure of model (\ref{full}) and especially the $\lambda_2$ term critically affects the spectrum of the 
 ${\mathbb{Z}}_4$ Fateev model (which in the given case is equivalent to the sine-Gordon model with $\beta^2 = 6\pi$
 (\ref{SGZ4})). Now we will try to discern the nature of the new excitations. 

 As we have said, the operator $\cos(\sqrt{2\pi/3}\chi)$ has two degenerate vacuum values.  When $\lambda_2 < 0$ the values $\chi = \pm \sqrt{\pi/6}$ correspond to vacua with $\Phi_a =0$. So one can suggest there are two types of kinks and antikinks of $\chi$-field: short kinks where $\sqrt{6\pi}\chi$ shifts by $2\pi$  and long kinks where it shifts by  $4\pi$. Both types of kinks are SU(4) neutral. 

  For $\lambda_2 > 0$ it is not clear what the vacuum configuration is. It may be the vacua $\chi = \pm \sqrt{\pi/6}$ and nonzero $\Phi_a$ or the single vacuum $\chi=0$. Which situation is realized is determined by the energetics.  In the first case there are short  and long kinks as before, in the second one there is only one type of kinks, the one where $\sqrt{6\pi}\chi$ shifts  by $6\pi$. 

 The second case corresponds to three particle confinement which has not been explored in the literature. This makes it worth commenting on which we do below. Averaging by small fluctuations of $\Phi_a$ fields in (\ref{full}) we obtain the effective theory in the form of two-frequency sime-Gordon model:
\bea
{\cal L}_{\rm para} = \frac{1}{2}(\p_{\mu}\chi)^2 - \lambda_1\cos(\sqrt{6\pi}\chi) - g\cos(\sqrt{2\pi/3}\chi).
\label{conflag}
\eea

 At $g=0$ the sine-Gordon model has only kinks and no breathers. When $g >0$ the situation changes. 
The second term in Eq. (\ref{conflag}) has three vacua: $\beta\chi = 0, 2\pi, -2\pi$. In the presence of the last term the last two vacua become false. This leads to confinement of $(0,2\pi)$ and $(0,-2\pi)$ kinks with formations of breathers (mesons). The kinks now are between 0 and $6\pi$ vacua (hadrons) and are formed by  confinement of three kinks  $(0,2\pi), (2\pi, 4\pi), (4\pi,6\pi)$. Both  mesons and hadrons  are SU(4) singlets.  Being topological excitations they will not experience decay from interaction with the SU(4) sector.

The confinement of two kinks can be described in a standard way. The spectrum of mesons starts from above $2M$ threshould, where $M$ is the kink's mass in the $\beta^2 = 6\pi$ sine-Gordon model. 

 The confinement of three kinks can be approximately described as follows. Only $(0,2\pi)$ and $(4\pi,6\pi)$ kinks interpolate between the vacua with different value of $\la \cos(\sqrt{2\pi/3}\chi)\ra$. Kinks $(2\pi,4\pi)$ do not change this expectation value. Therefore the linear potential exists only between the former kinks and the latter one can move freely in the space between the two. In the adiabatic approximation this movement contributes the amount 
\bea
\delta E = \frac{\pi^2}{2M|x_{12}|^2},
\eea
to the total energy. As a result the Schroedinger equation for the former kinks is 
\bea
\Big[-\frac{1}{2M}\p^2_1 -\frac{1}{2M}\p^2_2 + \eta |x_{12}| + \frac{\pi^2}{2M|x_{12}|^2}\Big]\Psi = (E- 3M)\Psi,
\eea
where $M$ is the kink's mass and $\eta \sim gM^{1/6}$ is the energy density of the false vacuum. 

\subsubsection{The SU(4) sector}

 Now let us return to the SU(4) sector.  The kinks  transform according to the fundamental representation of the SU(6) group, not the adjoint representation of the SU(4). 
 Apparently, the above vector bosons  are unstable and decay into the kinks, antikinks. 

 How to get the SU(4) out of this? It is possible that since one has to project out the center of mass mode (\ref{centermass}), the physical states are made of {\it confined} antisymmetric kink-antikink pairs which comprise 15-dimensional adjoint representation of the SU(4) group.  

 The easiest task is to calculate the breather contribution to the $\la g g^+\ra$ since the neutral breather can be made as a linear conbination of breathers from different sine-Gordon copies. Such excitation  just corresponds to the expansion of the cosines in (\ref{model2}) around their minima and hence does not contain the zero mode.  So the contribution is 
\bea
\frac{Z}{\omega^2 - p^2 - 3m^2}.
\eea
Since the kinks and antikinks appear in pairs, the energy  threshould for their emission is $2m$ or even higher if there is a real confinement and not just a condition of global neutrality. So it seems that the vector particles are indeed coherent as it has been  found before.

\section{Strong-coupling analysis of the lattice two-leg SU($2n$) spin ladder}

In this Appendix, we consider the standard two-leg SU($2n$) ladder model (\ref{ladder})  with $
\hat V = \hat P^{(1,2)}$ (SU($2n$) permutation operator between the two chains)  in the strong-coupling regime when $ \lambda  \rightarrow \infty$.
In that limit, the model is  equivalent to a single SU($2n$) spin chain where the spin operators belong to the antisymmetric representation defined by a Young tableau with a single column of two boxes. For $N=4$, the physics of the latter model is well understood and a dimerized phase with a two-fold ground-state degeneracy occurs. \cite{affleckmarston,marston,assarafPRL,marston07,nonne2011}
It corresponds to the plaquette phase of Fig. 1 found in our field theory analysis in the weak-coupling regime.

When $2n > 4$, the nature of the phase might be inferred by means of a non-Abelian bosonization approach starting
from the underlying U($2n$) Hubbard model at a filling of 2 atoms per site ($k_F = \pi/n a_0$). \cite{affleck}
At such a filling, there is an umklapp operator which couples the charge degrees of freedom with the non-Abelian
ones:
\begin{equation}
{\cal V}_{u} \sim e^{2 i n k_F x} \prod_{i=1}^{n} L^{\dagger}_{a_i} R_{a_i} + H.c. ,
\label{umklapp2atoms}
\end{equation}
where $L_{a_i}$ and $R_{a_i}$ are respectively the $2n$ left and right-moving Dirac fermions with $a_i = 1, \ldots, 2n$.
Using the non-Abelian bosonization rule similar to Eq. (\ref{grep}), we get a regular contribution:
\begin{equation}
{\cal V}_{u} \sim - g_u \cos \left({\sqrt{2 \pi n} \Phi_c}\right) :\left( {\rm Tr} g \right)^N:,
\label{umklappcont}
\end{equation}
where $\Phi_c$ is the charge bosonic field and $g$ is the SU(2n)$_1$ WZNW primary field with 
scaling dimension $1 - 1/2n$. The operator $:\left( {\rm Tr} g \right)^N:$ corresponds to the 
WZNW primary field with scaling dimension $n/2$ which transforms in the self-conjugate antisymmetric
representation of the SU($2n$) group with a Young tableau of a single column with $n$ boxes.  Assuming
that we have a charge gap $\Delta_c$ (with a Luttinger parameter $K_c < 4/n$), we have in the low-energy
limit $E \ll \Delta_c$ the effective interacting Hamiltonian for the non-Abelian degrees of freedom:
\begin{equation}
{\cal H}_{\rm eff} \simeq - g_u  :\left( {\rm Tr} g \right)^N:,
\label{umklappcontnonabel}
\end{equation}
which is a strongly relevant perturbation when $n \le 3$, marginal for $n=4$, and is irrelevant for $n \ge 5$.
A straightforward semiclassical analysis  gives rise, at $g_u>0$ and $n \le 3$, to an $n$-fold degenerate phase
which breaks spontaneously the one-step translation symmetry: $ g = e^{i k \pi/n } I$, $k = 0, \ldots, n-1$. In this
respect, a dimerized (respectively trimerized) phase is expected for $n=2$ (respectively, $n=3$), i.e., $N=4$
(respectively, $N=6$). In particular, the trimerized phase for $N=6$ where the spins form a 6-site rectangular
cluster, found in the weak-coupling regime, extends to the strong-coupling case.
Interestingly enough, when $n >4$, the interaction in Eq. (\ref{umklappcontnonabel}) becomes irrelevant
and an SU($2n$)$_1$ quantum criticality is then expected. For $n >4$, we thus a expect a quantum phase transition, in the 
SU($2n$)$_1$ universality class, for a finite value of the interchain coupling $\lambda$.
A recent variational Monte-carlo numerical analysis confirms these predictions and has shown that
the SU(8) case displays a quantum critical behavior. \cite{dufour}

\end{document}